# COSINE: A Web Server for <u>C</u>l<u>o</u>nal and <u>S</u>ubclonal Structure <u>In</u>ference and <u>E</u>volution in Cancer Genomics

Xiguo Yuan[1,#], Yuan Zhao[1,#], Yang Guo[1], Linmei Ge[2], Wei Liu[3], Shiyu Wen[3], Qi Li[1], Zhangbo Wan[1], Peina Zheng[1], Tao Guo[3], Zhida Li[3], Martin Peifer[4,*,a], Yupeng Cun[2,*,b]

[1]*School of Computer Science and Technology, Xidian University, Xi'an 710071, China*

[2]*iFlora Bioinformatics Center, Germplasm Bank of Wild Species, Kunming Institute of Botany, Chinese Academy of Sciences, Kunming 650201, China*

[3]*Yuxi Rongjian Information Technology Co., Ltd., Yuxi 653100, China*

[4]*Center for Molecular Medicine Cologne (CMMC), University of Cologne, Cologne 50931, Germany*

Symbol

[#] Equal contribution.

[*] Corresponding author(s).

E-mail: cunyupeng@mail.kib.ac.cn (Cun Y), mpeifer@uni-koeln.de (Peifer M)

**Running title:** *COSINE: Online Subclonal Structure Inference*

[a]ORCID: 0000-0002-5243-5503.

[b]ORCID: 0000-0002-4241-8099.

## Abstract

Cancers evolve from mutation of a single cell with sequential clonal and subclonal expansion of somatic mutation acquisition. Inferring clonal and subclonal structures from bulk or single cell tumor genomic sequencing data has a huge impact on cancer evolution studies. Clonal state and mutational order can provide detailed insight into tumor origin and its future development. In the past decade, a variety of methods have been developed for subclonal reconstruction using bulk tumor sequencing data. As these methods have been developed in different programming languages and using different input data formats, their use and comparison can be problematic. Therefore, we established a web server for clonal and subclonal structure inference and evolution of cancer genomic data (COSINE), which included 12 popular subclonal reconstruction methods. We decomposed each method via a detailed workflow of single processing steps with a user-friendly interface. To the best of our knowledge, this is the first web server providing online subclonal inference, including the most popular subclonal reconstruction methods. COSINE is freely accessible at www.clab-cosine.net or http://bio.rj.run:48996/cun-web.

## Introduction

The genome of cancer cells originate from mutation of a single cell with sequential clonal and subclonal expansion of somatic mutation acquisition during pathogenesis, which is thought to be a Darwinian evolutionary process [1-4]. Through next-generation sequencing (NGS) of tumor tissue, this evolutionary process can be characterized by statistical modelling, and the clonal state, somatic mutation order, and evolutionary processes can be identified [4-6]. Subclonal inference of overall tumor genome sequencing data is an important part of tumor evolution research, providing a new avenue for studying the relative order of mutations and the mutation process in tumorigenesis. This evolutionary process can be inferred from NGS data, with the assumption of "most recent common ancestor (MRCA)" adopted from classical population genetics.

In the past decade, a variety of subclonal reconstruction methods have been developed for large or single cell genomic data of single or multiple tumor samples over time and/or multiple locations [6-18]. The subclonal reconstruction process typically includes three steps: first, computing the variant allelic fractions of somatic mutations with related copy number alterations and tumor purity; second, estimating the cancer cell fraction (CCF) in the tumor, which using structural variant information for better accuracy; third, clustering the CCFs to identify subclonal structures and construct related phylogenetic trees. Through the process of clonal and subclonal expansion, the landscape of normal cells (i.e., common ancestors in population genetics) then evolves into different cancer cells. Thus, based on experimental design and reconstruction of specific tumor mutation characteristics, the accuracy and resolution of each feature inferred by subcloning can be determined. Among the above methods, most employ non-parametric Bayesian approaches (e.g., Dirichlet process with stick-break representation) for clustering [6, 8, 10, 14, 15], and employ Markov chain Monte Carlo (MCMC) resampling schemes that contain high computational costs, especially under high mutation rates. A more economical way for clustering is to use modified mixed Bayesian models, such as SciClone [9]. Combinational phylogeny is another popular approach used for clustering, and includes TrAP [7], CITUP [11], and CloneFinder [13], although CloneFinder only uses single nucleotide variant (SNV) information. The deconvolution of cancer cell SNV density shows a high computational efficiency for subclonal inferencing, as first applied in Sclust [12] and later in FastClone [17]. Both deconvolution methods can complete the subclonal inference process in less than 5 s using simulation data with more than 500 SNVs. Regularized maximum-likelihood estimation methods can also be used for subclonal inferencing, e.g., CLiP [16].

As subclonal reconstruction methods have been developed using different programming languages and are generally under the Linux platform, many users may find it difficult to operate and compare them. In this paper, we established a web server for clonal and subclonal structure inference and evolution in cancer genomics, which included 12 popular subclonal reconstruction methods [6-17], e.g., DPclust [6], PyClone [8], PhyloWGS [10], and Sclust [12]. Each method is decomposed through detailed operational steps and implemented through the relevant operational interface, which allows professional or non-professional bioinformatic scientists to run and compare methods using their own data. Although the comparison of some subclonal inference methods has been performed in this field [18], but online tools for subclonal inferencing remain scarce. To narrow the gap between model and user, we established the first web server to provide online subclonal inferencing, with the inclusion of the 12 most popular subclonal inferencing methods.

## Methods

### Subclonal inference from bulk genomic data

All tumor cells in a sample that mutate before the MRCA can be used as producers of a clonal population [4, 18]. Driver and passenger mutations continue to accumulate during tumor growth (Figure 1A). Tumor cells with driver mutations will trigger a clonal expansion, which creates a subpopulation of cells. Clonal mutation means all tumor cells bearing mutations, and subclonal mutations means only part of the tumor cells bearing mutations at a variant allele, and such a subclonal cells was identified through shared mutations. The clonal and subclonal mutations are shown in Figure 1B and 1C respectively. For example, in Figure 1B and 1C, there were 16 cells, which contain 10 tumor cells and six normal cells, in Figure 1B and 1C, so the purity of tumor cells was 10/16 = 0.625. The observed variant allele frequency (VAF) of clonal mutation was 10/32 = 0.315, and subclonal mutation was 5/32 = 0.15625; the expected VAF of clonal mutation was 10/32 = 0.315, and subclonal mutation was 5/16 = 0.15625. For the clonal mutation, the observed VAF was equal to or close to the expected VAF. For the subclonal mutation, the observed VAF was significantly less than the expected VAF.

A general workflow for clonal and subclonal structure inference usually includes five steps: 1) calling somatic mutation from tumor-normal matched NGS data; 2) calling gene copy number in NGS data; 3) estimating CCFs; 4) inferencing clonal and subclonal structure via clustering of CCFs; 5) constructing clonal and subclonal evolutionary trees. The subclonal inference workflow of the COSINE is depicted in Figure 2. These were a lots tools for mutation and copy number calling in cancer genomics, for example GATK [19] and VarScan2 [20]. Generally, during sequencing of tumor tissue, normal cells are mixed in the tumor tissue. Thus, normal tissue close to tumor tissues is needed to obtain an accurate measure of the VAF of mutated sites. Several studies have estimated tumor-cell purity and somatic mutations in tumor-only samples based on machine learning [21, 22], however tumor-only methods require high sequencing coverage. Under low sequencing coverage (i.e., sequencing depth < 30x), tumors with normal matched type genomic data are recommended for estimating tumor purity and somatic mutations. After getting somatic mutation and copy number change information, CCFs can be estimated. Different subclonal reconstruction methods have their CCF estimate strategies, which will lead to slightly difference in subclonal inference results [18]. Structure variation information will help to improve subclonal inferencing results, for example in the latest research of SVclone [15]. A phylogenetic tree can then be constructed after clustering of CCFs. Some of these 12 methods in the CONSINE can construct phylogenetic tree, for example PhyloWGS [10], FastClone [17], PhylogicNDT [14], CITUP [11], CloneFinder [13], TrAp [7]. Details on the functions of the 12 methods in COSINE are summarized in Table 1. Tarabichi et al.

provided a practical guide to subclonal inferencing from bulk cancer genomic sequencing data [23].

The clustering of CCFs and identification of clonal and subclonal mutations are the most important steps in subclonal inferencing. Non-parametric Bayesian methods are frequently employed in subclonal inferencing [6, 8, 10, 14, 15]. As somatic mutations in tumor cell populations are derived from unknown subclone numbers and with unknown CCFs distributions and clone/subclone(s) distributions. These unknown parameters of this inference process can be jointly estimated via a Bayesian Dirichlet model [24]. Given a set of observed somatic mutations, with the total read depth of each base and read number of variant alleles for each mutation, then:

$$y_i = Bin(N_i, \xi_i \pi_i), with\ \pi_i \sim DP(\alpha P_0),$$

where $y_i$ is the number of reads of the *i-th* mutation with reads $N_i$ and the expected fraction of reads $\xi_i$; $\alpha > 0$ is a scaling parameter; $DP()$ is a Dirichlet process function. This model can be considered as a stick-break representation Dirichlet process and can be applied for clustering via MCMC resampling and running at least 20 000 iterations, leading to a high computational cost, especially when SNV > 4 000.

To overcome highly computational cost of Bayesian Dirichlet type methods, a more economical clustering way were employed a variational Bayesian mixture model or combinational phylogenetic method, like SciClone, TrAP, CITUP, and CloneFinder. The most efficient way was directly deconvolution on the CCF density of all cancer cells showed a highly computational efficiency for subclonal inferencing, which used in the Sclust and the FastClone. Among these two ultra-fast subclonal inferencing methods, the Sclust can take additional structure variation information into account and jointly estimate copy number alteration, tumor-cell purity and subclonal structure in subclonal inferencing process. In one DREAM challenge's benchmark study on the evaluation of subclonal reconstruction methods [18], a regularized maximum-likelihood estimation method also shows an economical way to perform subclonal inferencing, such as CLiP, which only provides a Python package and short abstract for the method. The regularized maximum-likelihood could be another economical way for subclonal inferencing.

**Subclonal inferencing is a key step to understand cancer evolution**

The clonal and subclonal states of somatic mutations provide key information to understand intratumor heterogeneity, e.g., prognostic management, therapeutic strategy, and drug resistance [3, 4]. With the advance of computational models and longitudinal cancer genome studies, exploring the micro-evolutionary history of tumors has become more predictable, which should help improve our understanding of the role of the immune microenvironment in tumorigenesis. Thus, subclonal inferencing is a key step after mutation and gene copy number calling. An online web server for subclonal inferencing is needed to accelerate and enhance cancer studies.

**Prepared input data for subclonal reconstruction**

After mapping raw genomic sequencing data to the reference genome, filter and adjust the read by GATK standard pipeline [19] is needed. The mapping, mutational calling step is supper computational time cost and resource consuming, we made a practical guide for using the BWA-MEM package [25] to map raw clean reads to the reference genome, and then correcting the BAM file with the best practice of the GATK4. The GTAK4-corrected bam file was used for calling mutation and copy number alteration via VarScan2 package [20]. A detailed practical guide for mapping, mutational calling, copy number calling, and subclonal inferencing was described in the Supplementary Files.

**Design and implementation of the COSINE web server**

To facilitate the use of our previously developed Sclust approach and the 11 other methods, we developed an online subclonal inference web server called COSINE (freely available at www.clab-cosine.net). Among these 12 methods, seven use only one programming language (Sclust developed in C++; PyClone, FastClone, and CloneFinder developed in Python; DPclust, SciClone developed in R; TrAp developed in Java) and five use more than two programming languages. These methods are all run a Linux system, which can hinder non-professional users from achieving subclonal inferencing quickly. In the COSINE web server, we implemented 12 subclonal inference methods to the high-performance computing cluster. Users can use one to five commands to call the subclonal inference method directly in the method's frame box through their web interface, and then download the results of the operation after completion. Due to security settings, users are required to register and log in to perform online subclonal inferencing when completing large tasks. COSINE can be accessed free of charge from www.clab-cosine.net.

    A COSINE workflow is illustrated in Figure 3. We decomposed each subclonal reconstruction method using detailed running steps, and implemented a related running interface. Figure 3A uses pre-processed raw data to call SNVs and copy number (structure variation needed for some methods), as described in the Supplementary Files. Figure 3C shows the COSINE interface, and the function of the 12 methods are summarized in Table 1. As these input files for each subclonal reconstruction method differ from each other, we first divided each method into different running steps, and then created Python script to change the somatic mutation vcf and copy number alteration file to the format of each method (see in Supplementary File).

**COSINE web server usage**

In COSINE, we developed a user-friendly online computational platform for subclonal structure inferencing, as shown in Figure 3C. Users can follow the following steps for subclonal inferencing in COSINE: 1) Visit COSINE website and select (click) the desired method (Figure 3C-1); 2) Create a new task on the method page (Figure 3C-2); 3) Upload and run the program ((Figure 3C-2); 4) Download the results upon completion of task. We created a special page for users to post their issues when using the methods.

**Future developments**

We developed COSINE, an online computational platform for subclonal structure inferencing of the cancer genome, with integration of 12 popular subclonal inferencing methods for easy access and a user-friendly interface. Many subclonal inference models have been proposed in recent years, which has introduced issues for biomedical researchers regarding method choice, installation, and program running. COSINE fills the gap from model developer to normal user, making subclonal inferencing easier and more convenient. In the future, we will add additional functions and methods for online clonal evolutionary tree plotting and adjustment, and also include subclonal reconstruction methods from single cell genomic sequencing data.

**Authors' contributions**

YC, MP, and XY conceived and designed the study. YC, WL, SW, ZL, and TG developed the program and wrote the computer codes for the web server. YC, YZ, LG, YG, QL, ZW, and PZ designed the web interface. YC, YZ, and YG wrote the supplement practical guide. YC and XY wrote and edited the manuscript. All authors read and approved the final manuscript.

## Competing interests

The authors have declared no competing interests.

## Acknowledgments

This work was supported by the CAS Pioneer Hundred Talents Program and National Natural Science Function China fund (No. 32070683) awarded to YC.

# References

[1] Nowell P. The clonal evolution of tumor cell populations. Science 1976;194:23-8.

[2] Navin N, Kendall J, Troge J, Andrews P, Rodgers L, McIndoo J, et al. Tumour evolution inferred by single-cell sequencing. Nature 2011;472:90-4.

[3] McGranahan N, Swanton C. Clonal Heterogeneity and Tumor Evolution: Past, Present, and the Future. Cell 2017;168:613-28.

[4] Gerstung M, Jolly C, Leshchiner I, Dentro SC, Gonzalez S, Rosebrock D, et al. The evolutionary history of 2,658 cancers. Nature 2020;578:122-8.

[5] George J, Lim JS, Jang SJ, Cun Y, Ozretić L, Kong G, et al. Comprehensive genomic profiles of small cell lung cancer. Nature 2015;524:47-53.

[6] Nik-Zainal S, Van Loo P, Wedge David C, Alexandrov Ludmil B, Greenman Christopher D, Lau King W, et al. The Life History of 21 Breast Cancers. Cell 2012;149:994-1007.

[7] Strino F, Parisi F, Micsinai M, Kluger Y. TrAp: a tree approach for fingerprinting subclonal tumor composition. Nucleic Acids Research 2013;41:e165.

[8] Roth A, Khattra J, Yap D, Wan A, Laks E, Biele J, et al. PyClone: statistical inference of clonal population structure in cancer. Nature Methods 2014;11:396-8.

[9] Miller CA, White BS, Dees ND, Griffith M, Welch JS, Griffith OL, et al. SciClone: Inferring Clonal Architecture and Tracking the Spatial and Temporal Patterns of Tumor Evolution. PLOS Computational Biology 2014;10:e1003665.

[10] Deshwar AG, Vembu S, Yung CK, Jang GH, Stein L, Morris Q. PhyloWGS: Reconstructing subclonal composition and evolution from whole-genome sequencing of tumors. Genome Biology 2015;16:35.

[11] Malikic S, McPherson AW, Donmez N, Sahinalp CS. Clonality inference in multiple tumor samples using phylogeny. Bioinformatics 2015;31:1349-56.

[12] Cun Y, Yang TP, Achter V, Lang U, Peifer M. Copy-number analysis and inference of subclonal populations in cancer genomes using Sclust. Nat Protoc 2018;13:1488-501.

[13] Miura S, Gomez K, Murillo O, Huuki LA, Vu T, Buturla T, et al. Predicting clone genotypes from tumor bulk sequencing of multiple samples. Bioinformatics 2018;34:4017-26.

[14] Gruber M, Bozic I, Leshchiner I, Livitz D, Stevenson K, Rassenti L, et al. Growth dynamics in naturally progressing chronic lymphocytic leukaemia. Nature 2019;570:474-9.

[15] Cmero M, Yuan K, Ong CS, Schröder J, Adams DJ, Anur P, et al. Inferring structural variant cancer cell fraction. Nature Communications 2020;11:730.

[16] Jiang Y, Yu K, Zhu H, Wang W. Abstract PO-029: CliP: A model-based method for subclonal architecture reconstruction using regularized maximum likelihood estimation. 2020.

[17] Xiao Y, Wang X, Zhang H, Ulintz PJ, Li H, Guan Y. FastClone is a probabilistic tool for deconvoluting tumor heterogeneity in bulk-sequencing samples. Nature Communications 2020;11:4469.

[18] Salcedo A, Tarabichi M, Espiritu SMG, Deshwar AG, David M, Wilson NM, et al. A community effort to create standards for evaluating tumor subclonal reconstruction. Nature Biotechnology 2020;38:97-107.

[19] McKenna A, Hanna M, Banks E, Sivachenko A, Cibulskis K, Kernytsky A, et al. The Genome Analysis Toolkit: a MapReduce framework for analyzing next-generation DNA sequencing data. Genome Res 2010;20:1297-303.

[20] Koboldt D, Zhang Q, Larson D, Shen D, McLellan M, Lin L, et al. VarScan 2: Somatic mutation and copy number alteration discovery in cancer by exome sequencing. Genome research 2012;22:568-76.

[21] Mao YF, Yuan XG, Cun YP. A novel machine learning approach (svmSomatic) to distinguish somatic and germline mutations using next-generation sequencing data. Zool Res 2021:1-4.

[22] Kalatskaya I, Trinh QM, Spears M, McPherson JD, Bartlett JMS, Stein L. ISOWN: accurate somatic mutation identification in the absence of normal tissue controls. Genome Med 2017;9:59.

[23] Tarabichi M, Salcedo A, Deshwar AG, Ni Leathlobhair M, Wintersinger J, Wedge DC, et al. A practical guide to cancer subclonal reconstruction from DNA sequencing. Nature Methods 2021;18:144-55.

[24] Dunson DB. Nonparametric Bayes applications to biostatistics. In: Holmes C., Hjort N. L., Müller P., Walker S. G. eds). Bayesian Nonparametrics. Cambridge: Cambridge University Press, 2010, 223-73.

[25] Li H. Aligning sequence reads, clone sequences and assembly contigs with BWA-MEM. arXiv:1303.3997v2 [q-bio.GN]. 2013.

# Figure legends

**Figure 1 Clonal expansion and related clonal and subclonal mutations.**

(A) Dynamic clonal expansion during cancer evolution. (B) Clonal type mutation, i.e., all tumor cells contain this somatic mutation at a specific site. (C) Subclonal type mutation, i.e., not all tumor cells contain this somatic mutation at a specific site.

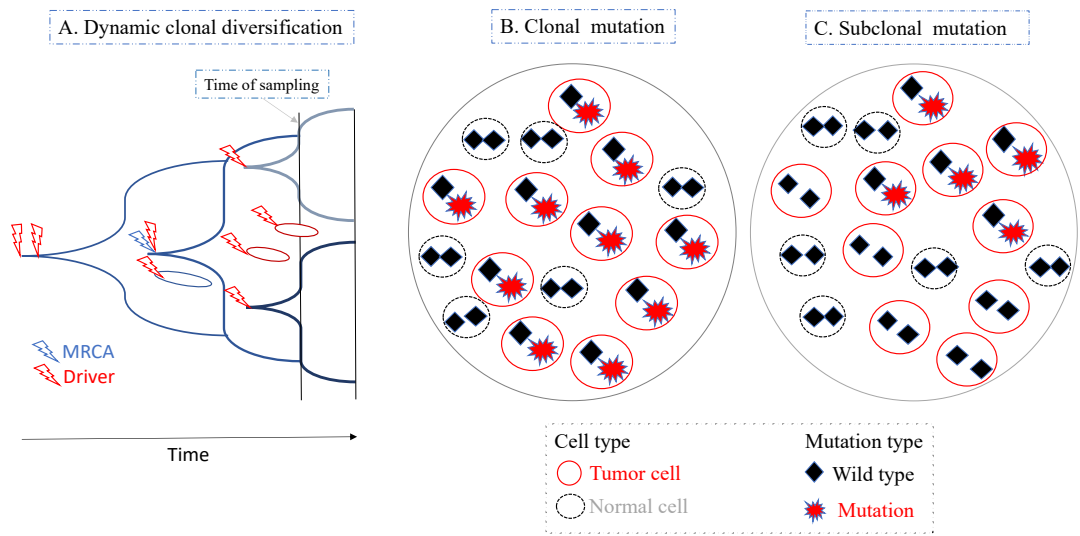

**Figure 2 General workflow for subclonal inferencing of bulk cancer genomic data**

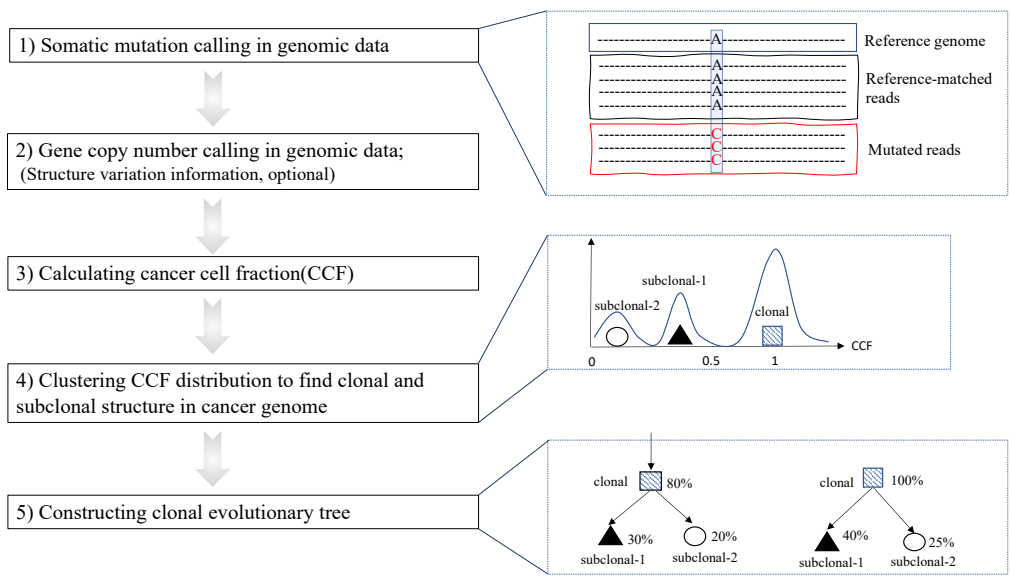

**Figure 3 Workflow of COSINE web server**

**(A) and (B)**. Typical pipeline for mapping, bam filtering, mutation and copy number/structure variation calling of raw clean data. Somatic mutation vcf and copy number alteration and structure variation (optimal) information were used for subclonal reconstruction. **(C)**. Twelve subclonal reconstruction methods with detailed step(s) in COSINE.

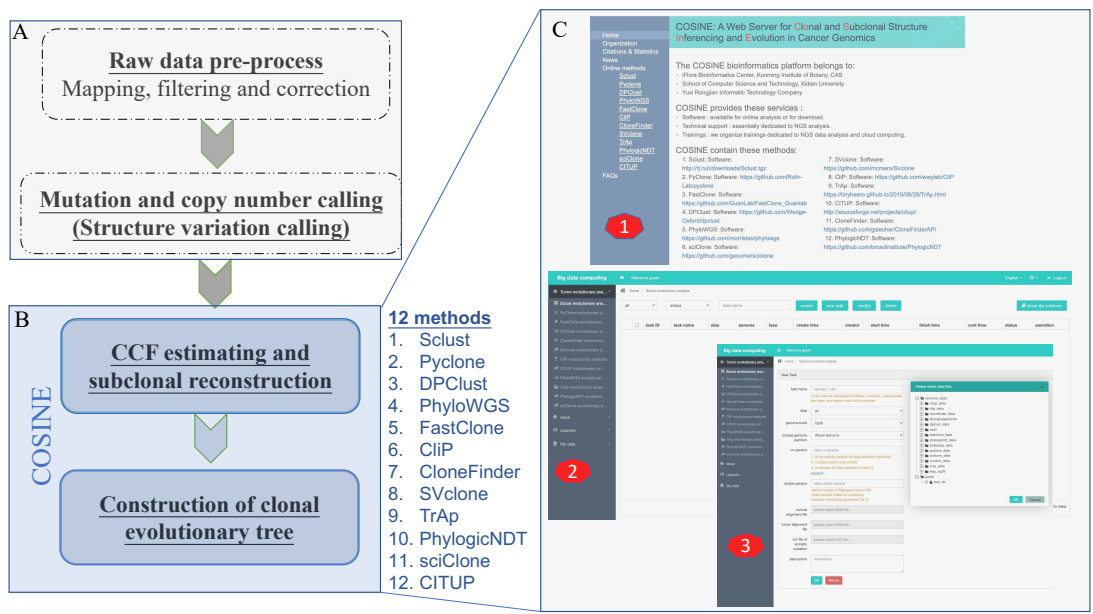

# Tables

## Table 1 Comparison of copy-number and subclonal architecture inference methods

| Property | Sclust | SVclone | PhyloWGS | FastClone | PyClone | DPClust | PhylogicNDT | CITUP | SciClone | CliP | CloneFinder | TrAp |
|---|---|---|---|---|---|---|---|---|---|---|---|---|
| Performs own copy-number segmentation | Y | N | N | N | N | N | N | N | N | N | N | N |
| Uses rearrangement breakpoints in segmentations | Y | Y | N | N | N | N | N | N | N | N | N | N |
| Calls absolute clonal copy numbers | Y | Y | N | N | N | N | N | N | N | N | N | N |
| Calls subclonal copy numbers | Y | Y | N | N | N | N | N | N | N | N | N | N |
| Clusters copy number alteration | Y | Y | Y | N | N | N | N | N | N | N | N | N |
| Deals with sample(s) | 1 | 1 | ≥1 | 1 | ≥1 | ≥1 | ≥1 | ≥1 | ≥1 | 1 | ≥1 | ≥1 |
| Reconstructs phylogenetic trees | N | N | Y | Y | N | N | Y | Y | N | N | Y | Y |
| Allows subclonal copy numbers for clustering | Y | Y | Y | Y | Y | Y | N | N | Y | N | N | N |
| Programming language | C++ | R Python | Python Tex JavaScript, C++ HTML CSS | Python | Python | R | HTML Python | C++ Python | R | R Python | Python | Java |

"Y" denotes the method performs the corresponding function.

# Supplementary material